\documentclass[twocolumn,showpacs,aps,prl,superscriptaddress]{revtex4}

\usepackage{graphicx}
\usepackage{dcolumn}
\usepackage{amsmath}
\usepackage{epsfig}

\RequirePackage{xspace}

\usepackage{relsize}
\def\babar{\mbox{\slshape B\kern-0.1em{\smaller A}\kern-0.1em
    B\kern-0.1em{\smaller A\kern-0.2em R}}}

\def\epem       {\ensuremath{e^+e^-}\xspace}

\def\qqbar {\ensuremath{q\overline q}\xspace}
\def\u     {\ensuremath{u}\xspace}
\def\ubar  {\ensuremath{\overline u}\xspace}

\def\d     {\ensuremath{d}\xspace}

\def\b     {\ensuremath{b}\xspace}

\def\piz   {\ensuremath{\pi^0}\xspace}

\def\pim   {\ensuremath{\pi^-}\xspace}

\def\Kbar  {\kern 0.2em\overline{\kern -0.2em K}{}\xspace}

\def\Kz    {\ensuremath{K^0}\xspace}
\def\Kzb   {\ensuremath{\Kbar^0}\xspace}
\def\KzKzb {\ensuremath{\Kz \kern -0.16em \Kzb}\xspace}
\def\Kp    {\ensuremath{K^+}\xspace}
\def\Km    {\ensuremath{K^-}\xspace}

\def\KpKm  {\ensuremath{\Kp \kern -0.16em \Km}\xspace}

\def\Dbar    {\kern 0.2em\overline{\kern -0.2em D}{}\xspace}

\def\Dz      {\ensuremath{D^0}\xspace}
\def\Dzb     {\ensuremath{\Dbar^0}\xspace}
\def\DzDzb   {\ensuremath{\Dz {\kern -0.16em \Dzb}}\xspace}
\def\Dp      {\ensuremath{D^+}\xspace}
\def\Dm      {\ensuremath{D^-}\xspace}

\def\DpDm    {\ensuremath{\Dp {\kern -0.16em \Dm}}\xspace}

\def\B       {\ensuremath{B}\xspace}
\def\Bbar    {\kern 0.18em\overline{\kern -0.18em B}{}\xspace}

\def\BB      {\ensuremath{B\Bbar}\xspace} 
\def\Bz      {\ensuremath{B^0}\xspace}
\def\Bzb     {\ensuremath{\Bbar^0}\xspace}
\def\BzBzb   {\ensuremath{\Bz {\kern -0.16em \Bzb}}\xspace}
\def\Bu      {\ensuremath{B^+}\xspace}
\def\Bub     {\ensuremath{B^-}\xspace}
\def\Bp      {\ensuremath{\Bu}\xspace}

\def\Bpm     {\ensuremath{B^\pm}\xspace}

\def\BpBm    {\ensuremath{\Bu {\kern -0.16em \Bub}}\xspace}

\def\BorBbar    {\kern 0.18em\optbar{\kern -0.18em B}{}\xspace}
\def\DorDbar    {\kern 0.18em\optbar{\kern -0.18em D}{}\xspace}
\def\KorKbar    {\kern 0.18em\optbar{\kern -0.18em K}{}\xspace}

\mathchardef\Upsilon="7107
\def\Y#1S{\ensuremath{\Upsilon{(#1S)}}\xspace}

\def\FourS {\Y4S}

\mathchardef\Deltares="7101
\mathchardef\Xi="7104
\mathchardef\Lambda="7103
\mathchardef\Sigma="7106
\mathchardef\Omega="710A

\def\Deltabar{\kern 0.25em\overline{\kern -0.25em \Deltares}{}\xspace}
\def\Lbar{\kern 0.2em\overline{\kern -0.2em\Lambda\kern 0.05em}\kern-0.05em{}\xspace}
\def\Sigbar{\kern 0.2em\overline{\kern -0.2em \Sigma}{}\xspace}
\def\Xibar{\kern 0.2em\overline{\kern -0.2em \Xi}{}\xspace}
\def\Obar{\kern 0.2em\overline{\kern -0.2em \Omega}{}\xspace}
\def\Nbar{\kern 0.2em\overline{\kern -0.2em N}{}\xspace}
\def\Xb{\kern 0.2em\overline{\kern -0.2em X}{}\xspace}

\def\Bztorhopi  {\ensuremath{\Bz \to \rho^+ \pim}\xspace}

\def\upsbb   {\ensuremath{\FourS \to \BB}\xspace}

\def\mes        {\mbox{$m_{\rm ES}$}\xspace}

\def\DeltaE     {\mbox{$\Delta E$}\xspace}

\newcommand{\tev}{\ensuremath{\mathrm{\,Te\kern -0.1em V}}\xspace}
\newcommand{\gev}{\ensuremath{\mathrm{\,Ge\kern -0.1em V}}\xspace}
\newcommand{\mev}{\ensuremath{\mathrm{\,Me\kern -0.1em V}}\xspace}
\newcommand{\kev}{\ensuremath{\mathrm{\,ke\kern -0.1em V}}\xspace}
\newcommand{\ev}{\ensuremath{\mathrm{\,e\kern -0.1em V}}\xspace}
\newcommand{\gevc}{\ensuremath{{\mathrm{\,Ge\kern -0.1em V\!/}c}}\xspace}
\newcommand{\mevc}{\ensuremath{{\mathrm{\,Me\kern -0.1em V\!/}c}}\xspace}
\newcommand{\gevcc}{\ensuremath{{\mathrm{\,Ge\kern -0.1em V\!/}c^2}}\xspace}
\newcommand{\mevcc}{\ensuremath{{\mathrm{\,Me\kern -0.1em V\!/}c^2}}\xspace}

\def\mus  {\ensuremath{\rm \,\mus}\xspace}

\def\ps   {\ensuremath{\rm \,ps}\xspace}

\def\mus        {\ensuremath{\,\mu{\rm s}}\xspace}    
\def\ps         {\ensuremath{{\rm \,ps}}\xspace}  

\def\to                 {\ensuremath{\rightarrow}\xspace}

\newcommand{\stat}{\ensuremath{\mathrm{(stat)}}\xspace}
\newcommand{\syst}{\ensuremath{\mathrm{(syst)}}\xspace}

\def\pep2{PEP-II}
\def\BF{$B$ Factory}
\def\abf {asymmetric \BF}

\def\gsim{{~\raise.15em\hbox{$>$}\kern-.85em
          \lower.35em\hbox{$\sim$}~}\xspace}
\def\lsim{{~\raise.15em\hbox{$<$}\kern-.85em
          \lower.35em\hbox{$\sim$}~}\xspace}

\def\CP                {\ensuremath{C\!P}\xspace}
\def\C       {\ensuremath{C}\xspace}

\def\deltaz{\ensuremath{{\rm \Delta}z}\xspace}
\def\deltat{\ensuremath{{\rm \Delta}t}\xspace}
\def\deltamd{\ensuremath{{\rm \Delta}m_d}\xspace}

\xspace

\newcommand{\jprlBase}       {Phys.\ Rev.\ Lett.\xspace}
\newcommand{\jprBase}        {Phys.\ Rev.\xspace}
\newcommand{\jplBase}        {Phys.\ Lett.\xspace}
\newcommand{\nimBaseA}       {Nucl.\ Instr.\ Meth.\xspace}

\newcommand{\cpc}       [1]  {{Comput.\ Phys.\ Commun.\ {\bf #1}}}

\newcommand{\nima}      [1]  {\nimBaseA~A~{\bf #1}}

\newcommand{\plb}       [1]  {\jplBase\ B~{\bf #1}}

\newcommand{\jprl}      [1]  {\jprlBase\ {\bf #1}}
\newcommand{\jprd}      [1]  {\jprBase\ D~{\bf #1}}

\newcommand{\progtp}    [1]  {{Prog.\ Theor.\ Phys.\ {\bf #1}}}

\def\jetset74   {\mbox{\tt Jetset \hspace{-0.5em}7.\hspace{-0.2em}4}\xspace}

\newcommand{\BABARPubYear}    {04}
\newcommand{\BABARPubNumber}  {09}

\newcommand{\SLACPubNumber} {10399}

\def\Bztorhoprhom {\ensuremath{\Bz (\Bzb) \to \rho^+ \rho^- }\xspace}
\def\Bztopippim {\ensuremath{\Bz (\Bzb) \to \pi^+ \pi^- }\xspace}
\def\rhoprhom {\ensuremath{\rho^+ \rho^- }\xspace}

\def\clong   {\ensuremath{ C_{L} }}
\def\slong   {\ensuremath{ S_{L} }}
\def\ct   {\ensuremath{ C_{T} }}
\def\st   {\ensuremath{ S_{T} }}
\def\fL   {\ensuremath{ f_L }}
\def\ptrue   { \fL }

\def\coshel  {\ensuremath{ \cos\theta_{i} }}
\def\mv      {\ensuremath{ m_{\pi^\pm \pi^0 }}}

\def\nno     {\ensuremath{\cal{N}}}

\def\figurebox#1#2#3{%
    \def\arg{#3}%
    \ifx\arg\empty
    {\hfill\vbox{\hsize#2\hrule\hbox to #2{\vrule\hfill\vbox to #1{\hsize#2\vfill}\vrule}\hrule}\hfill}%
    \else
    {\hfill\epsfbox{#3}\hfill}%
    \fi}

\begin{document}

\preprint{\babar-PUB-\BABARPubYear/\BABARPubNumber}
\preprint{SLAC-PUB-\SLACPubNumber}

\begin{flushleft}
\babar-PUB-\BABARPubYear/\BABARPubNumber\\
SLAC-PUB-\SLACPubNumber\\
\end{flushleft}

\title{
{\large \bf Study of the decay \Bztorhoprhom, and constraints on the CKM angle $\alpha$.}}

%
\author{B.~Aubert}
\author{R.~Barate}
\author{D.~Boutigny}
\author{F.~Couderc}
\author{J.-M.~Gaillard}
\author{A.~Hicheur}
\author{Y.~Karyotakis}
\author{J.~P.~Lees}
\author{V.~Tisserand}
\author{A.~Zghiche}
\affiliation{Laboratoire de Physique des Particules, F-74941 Annecy-le-Vieux, France }
% ...
\collaboration{The \babar\ Collaboration}
\noaffiliation

\date{\today}

\begin{abstract}

%
%
Using a data sample of 89 million \upsbb decays collected with the \babar\ detector at the
\pep2\ \abf\ at SLAC, we measure the  \Bztorhoprhom\ branching fraction as $(30
\pm 4 \stat \pm 5 \syst) \times 10^{-6}$ and a longitudinal polarization fraction of
$\ptrue = 0.99 \pm 0.03 \stat \;^{+0.04}_{-0.03} \syst$. We measure the 
time-dependent-asymmetry parameters of the longitudinally polarized component of 
this decay as $\clong   = -0.17\pm 0.27 \stat \pm 0.14 \syst$ and 
${\slong} =  -0.42 \pm 0.42 \stat \pm 0.14 \syst$.  We exclude values of 
$\alpha$ between $19^\circ$ and $71^\circ$ (90\% C.L.).
\end{abstract}


\maketitle

\maketitle
%
%

The recently observed~\cite{LBLrr} decay \Bztorhoprhom proceeds mainly through
the $\b \to \u \ubar \d$ tree diagram. Interference between direct decay and decay after 
\Bz-\Bzb mixing results in a time-dependent decay-rate asymmetry between
\Bz and \Bzb that is sensitive to the CKM~\cite{CKM} angle $\alpha \equiv
\arg\left[-V_{td}^{}V_{tb}^{*}/V_{ud}^{}V_{ub}^{*}\right]$. The presence of 
loop (penguin) contributions introduces additional phases that can shift the 
experimentally measurable parameter $\alpha_{\mathrm{eff}}$ away from the value 
of $\alpha$.  In the presence of penguin contributions
$\alpha_{\mathrm{eff}} = \alpha + \delta\alpha_{\mathrm{penguin}}$.
A constraint on $\alpha$ tests the Standard Model description 
of CP violation.
Recent measurements of the $\Bp \to \rho^+\rho^0$ branching fraction and upper limit for
$\Bz \to \rho^0 \rho^0$ \cite{recentrhorho} indicate small penguin contributions in 
$\B \to \rho \rho$, as has been found in some calculations~\cite{aleksan}.
Here we present a time-dependent analysis of \Bztorhoprhom.

The \CP analysis of $B$ decays to $\rho^+\rho^-$ is complicated by the presence of three
helicity states ($h=0,\pm 1$). The $h=0$ state corresponds to longitudinal polarization and
is \CP-even, while neither the $h=+1$ nor the $h=-1$ state is an eigenstate of \CP.
The longitudinal polarization fraction $f_L$ is defined as the fraction of
the helicity zero state in the decay. The angular distribution is 
\begin{eqnarray}
&&\frac{d^2\Gamma}{\Gamma d\cos\theta_1 d\cos\theta_2
}= \nonumber \\
&&\frac{9}{4}\left(f_L \cos^2\theta_1 \cos^2\theta_2 + \frac{1}{4}(1-f_L) \sin^2\theta_1 \sin^2\theta_2 \right)
\end{eqnarray}
where $\theta_{i}, i=1,2$ is defined for each $\rho$ meson as the angle 
between the \piz momentum in the $\rho$ rest frame and the flight direction of the $B^0$ 
in this frame. We have integrated over the angle between the $\rho$-decay planes. 
A full angular analysis of the decays is needed in order to separate
the definite \CP contributions; if however a single \CP channel dominates the decay, 
this is not necessary~\cite{Dunietz}.  The longitudinal polarization dominates this 
decay~\cite{Suzuki,LBLrr}.

This measurement is based on 89 million \upsbb decays collected
with the \babar ~\cite{babar} detector at the \pep2\ \abf\ at SLAC.
We reconstruct \Bztorhoprhom candidates ($B_{\rm rec}$) from combinations of two charged
tracks and two \piz candidates. We require that both tracks have particle identification
information inconsistent with the electron, kaon, and proton hypotheses. The \piz candidates
are formed from pairs of photons that have measured energies greater than $50~\mev$. The
reconstructed \piz\ mass must satisfy $0.10 < m_{\gamma\gamma} < 0.16~\gevcc$. The mass of the $\rho$
candidates, \mv, must satisfy $|\mv-0.770~\gevcc | <0.375~\gevcc$. When multiple \B candidates can
be formed we select the one that minimizes the sum of the deviations of the
reconstructed \piz masses from the true \piz mass.
Combinatorial backgrounds dominate near $|\coshel|=1$,
while backgrounds from $B$ decays, like \Bztorhopi, with an additional low energy \piz from
the rest of the event (ROE), tend to concentrate at negative values of $\coshel$. We
reduce these backgrounds with the requirement $-0.8 < \coshel < 0.98$.

Continuum $\epem \to \qqbar$ ($q = u,d,s,c$) events are the dominant background.  
To discriminate signal from continuum we use a neural network 
(\nno) to combine six variables: the two event-shape variables used 
in the Fisher discriminant of Ref.~\cite{pipiBabar}; the cosine of the angle between 
the direction of the \B and the collision axis ($z$) in the
center-of-mass (CM) frame; the cosine of the angle between the \B thrust axis and the $z$ axis;
the cosine of the angle between the \B thrust axis and the thrust axis of the ROE; 
the decay angle of the \piz (defined in analogy to the $\rho$ decay angle, $\theta_i$);
the sum of transverse momenta in the ROE relative to the $z$ axis.

Signal events are identified kinematically using two variables, the difference \DeltaE
between the CM energy of the \B candidate and $\sqrt{s}/2$, and the
beam-energy substituted mass $\mes = \sqrt{(s/2 + {\mathbf {p}}_i\cdot {\mathbf
{p}}_B)^2/E_i^2- {\mathbf {p}}_B^2}$, where $\sqrt{s}$ is the total CM energy. The \B
momentum ${\mathbf {p}_B}$ and four-momentum of the initial state $(E_i, {\mathbf
{p}_i})$ are defined in the laboratory frame. We accept candidates that satisfy $5.21 < \mes
<5.29~\gevcc$ and $-0.12<\DeltaE<0.15~\gev$. The asymmetric \DeltaE window suppresses
background from higher-multiplicity \B decays.

To study the time-dependent asymmetry one needs to 
measure the proper time difference, \deltat, between the two \B\ decays 
in the event, and to determine the flavor tag of the 
other \B-meson.
The time difference between the decays of the two neutral
$B$ mesons in the event ($B_{\rm rec}$, $B_{\rm tag}$)
is calculated from the measured separation
\deltaz between the $B_{\rm rec}$ and $B_{\rm tag}$ decay
vertices~\cite{prdsin2b,BaBarSin2beta2}. We determine the $B_{\rm rec}$ vertex
from the two charged-pion tracks in its decay. The $B_{\rm tag}$
decay vertex is obtained by fitting the other tracks in the event,
with constraints from the $B_{\rm rec}$ momentum and the 
beam-spot location. The RMS resolution on
$\deltat$ is 1.1 \ps. We only use events for which the proper time
difference between the $B_{\rm rec}$ and $B_{\rm tag}$ decays
satisfies $|\deltat|<20 \, \ps$ and the error on \deltat, $\sigma(\deltat)$, is
less than $2.5 \, \ps$.
The flavor of the $B_{\rm tag}$ meson is determined with a multivariate
technique~\cite{pipiBabar} that has a total effective tagging efficiency of
$(28.4\pm0.7)$\%.  The events are assigned to five mutually exclusive taggging
categories {\tt Lepton}, {\tt Kaon 1}, {\tt Kaon 2}, {\tt Inclusive}, and {\tt Untagged},
listed in order of decreasing reliability of the tag.

Signal candidates may pass the selection even if one or more of the pions assigned to 
the $\rho^+\rho^-$ state is from the other $B$ in the event. 
These self-cross-feed (SCF) candidates comprise 39\% (16\%) of the accepted signal 
for $\fL=1$ ($\fL=0$).

The efficiency of the selection is 7.7\% (14.9\%) for longitudinally (transversely)
polarized signal as determined with Monte Carlo (MC)~\cite{geant}. 
The signal efficiency taking into account the measured 
polarisation is 7.7\%. We select 24288 events, which are dominated by
combinatoric backgrounds: roughly $86$\% from \qqbar and $13$\% from \BB. We distinguish
the following candidate types: (i) correctly reconstructed signal, for \Bztorhoprhom decays
where the correct particles are combined to form the $B_{\rm rec}$ candidate; (ii) SCF
signal; (iii) charm $\Bpm$ background ($b\to c$); (iv) charm $\Bz$ background ($b\to c$);
(v) charmless $B$ backgrounds; (vi) continuum $\epem \to \qqbar$ ($q = u,d,s,c$)
background. We consider both types (i) and (ii) as signal. 
The charmless decays
$\Bpm \to \rho^{\pm}\piz$, $\Bpm \to \rho^{\pm}\rho^0$, $\Bpm \to a_1^\pm\piz$, and $\B^{\pm, 0} \to
a_1\rho$ are expected to contribute to the final sample. For these decays we assume the following 
branching fractions: $11.0\pm 2.7$~\cite{babarrhopizero}, 
$26.4^{+6.1}_{-6.4}$~\cite{hfagrhorhozero}, $30\pm 15$, and $20\pm 20$, in units of 
$10^{-6}$, corresponding to $17\pm 4$, $16 \pm 4$, $30\pm 15$, 
and $26\pm 26$ events in the data, respectively.
The latter two are estimated from the measured branching 
fractions of related decays. We expect an additional $283 \pm 283$
candidates of charmless \B decays with more than four mesons in the final state; 
since most branching fractions for such modes have not been measured yet, we
generate them using the JETSET simulation~\cite{jetsetmc}. We expect 1700 (1016) 
charged (neutral) \B\ decays to final states containing charm mesons.
The rest of the background is composed of continuum.  Each of these backgrounds 
is included as an individual component in the fit, where the continuum yield is 
allowed to vary in the fit.

Each candidate is described with the eight $B_{\rm rec}$ kinematic 
variables \mes and \DeltaE, the \mv\ and \coshel\ values of the two $\rho$ 
mesons, \deltat, and \nno.  For each different candidate-type considered, 
we construct a probability density function (PDF) that is the product of PDFs
in each of these variables, assuming that they are uncorrelated.  The
total PDF is used in the fit to data. 

The parameters of the PDFs for continuum-background \mes, \DeltaE, \coshel, and 
\nno are allowed to vary in the final
fit to the data. The distribution of the continuum as a function of \mv\ is described by 
a non-parametric PDF~\cite{keys} derived from 
 \mes\ and \DeltaE\ data sidebands. For all other types these
distributions are extracted from high-statistics MC samples. The \coshel\ distributions for
the background are described by a non-parametric PDF derived from the MC, as the detector
acceptance and selection criteria modify the known vector-meson decay distribution. The
signal distribution is given by Eq. (1) multiplied by an acceptance function 
determined from signal MC.  We take into account known differences between data and the MC.

The signal \deltat\ distribution is described by an exponential (\B lifetime) multiplied by a
 \CP violating term, convoluted with three Gaussians ($\sim 90\%$ core, $\sim 9\%$ 
tail, $\sim 1\%$ outliers) and takes into 
account $\sigma(\deltat)$ from the vertex fit. The resolution is parameterized using a
large sample of fully reconstructed hadronic \B decays~\cite{prdsin2b}. 
 The nominal \deltat\ distribution for the \B
backgrounds is a non-parametric representation of the MC; in the study of systematic errors
we replace this model with the one used for signal. 
The continuum background is described by the sum of three Gaussian distributions whose 
parameters are determined by fitting the data.  

The signal decay-rate distribution 
$f_+ (f_-)$ for $B_{\rm tag}$= \Bz (\Bzb) is given by:
\begin{eqnarray*}
f_{\pm}(\deltat) = \frac{e^{-\left|\deltat\right|/\tau}}{4\tau} [1
\pm S\sin(\deltamd\deltat) \mp \C\cos(\deltamd\deltat)]\,,  \nonumber
\end{eqnarray*}
where $\tau$ is the mean \Bz lifetime, \deltamd is the 
\Bz-\Bzb mixing frequency, and 
$S$= \slong\ or \st\ and $C$= \clong\ or \ct\ are the \CP asymmetry parameters 
for the longitudinal and transversely polarized signal. 
The fitting function takes into account mistag dilution and is convoluted with the
\deltat resolution function described above. 
We set \st\ and \ct\ to zero since the transverse polarisation in the fit is small.

\begin{figure}[tp]
\begin{center}
\resizebox{8cm}{!}{
 \includegraphics{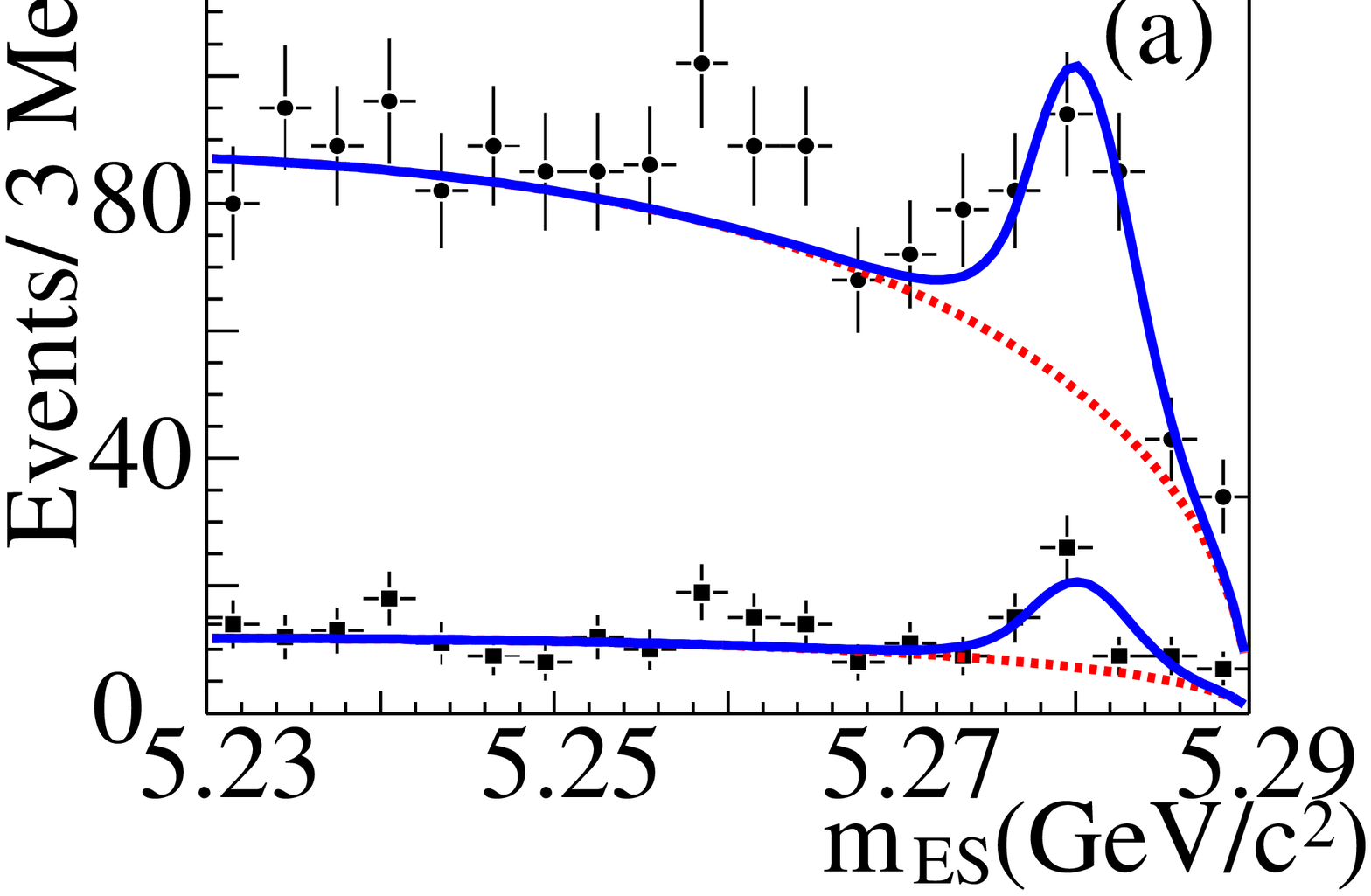}
 \includegraphics{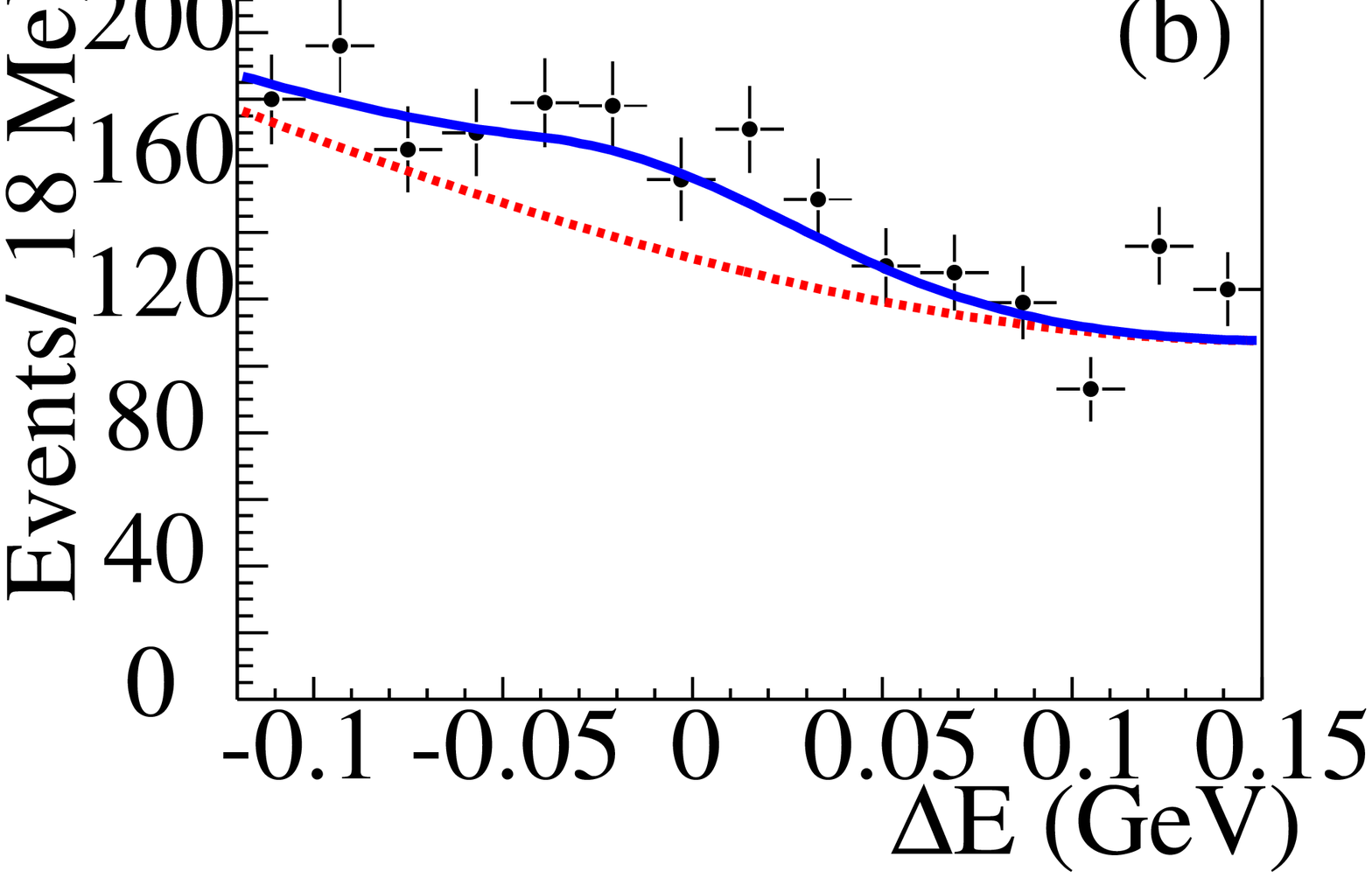}
}
\resizebox{8cm}{!}{
 \includegraphics{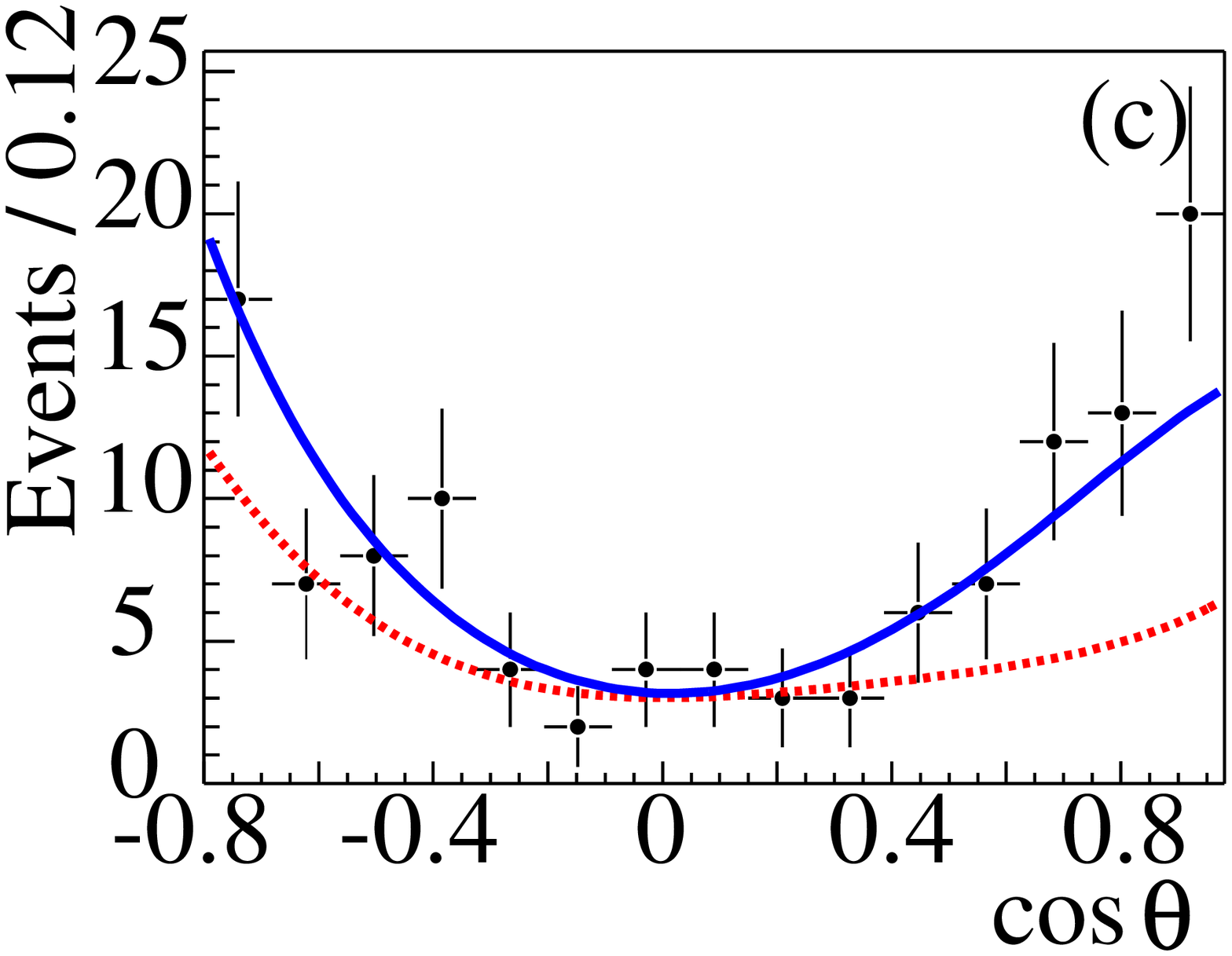}
 \includegraphics{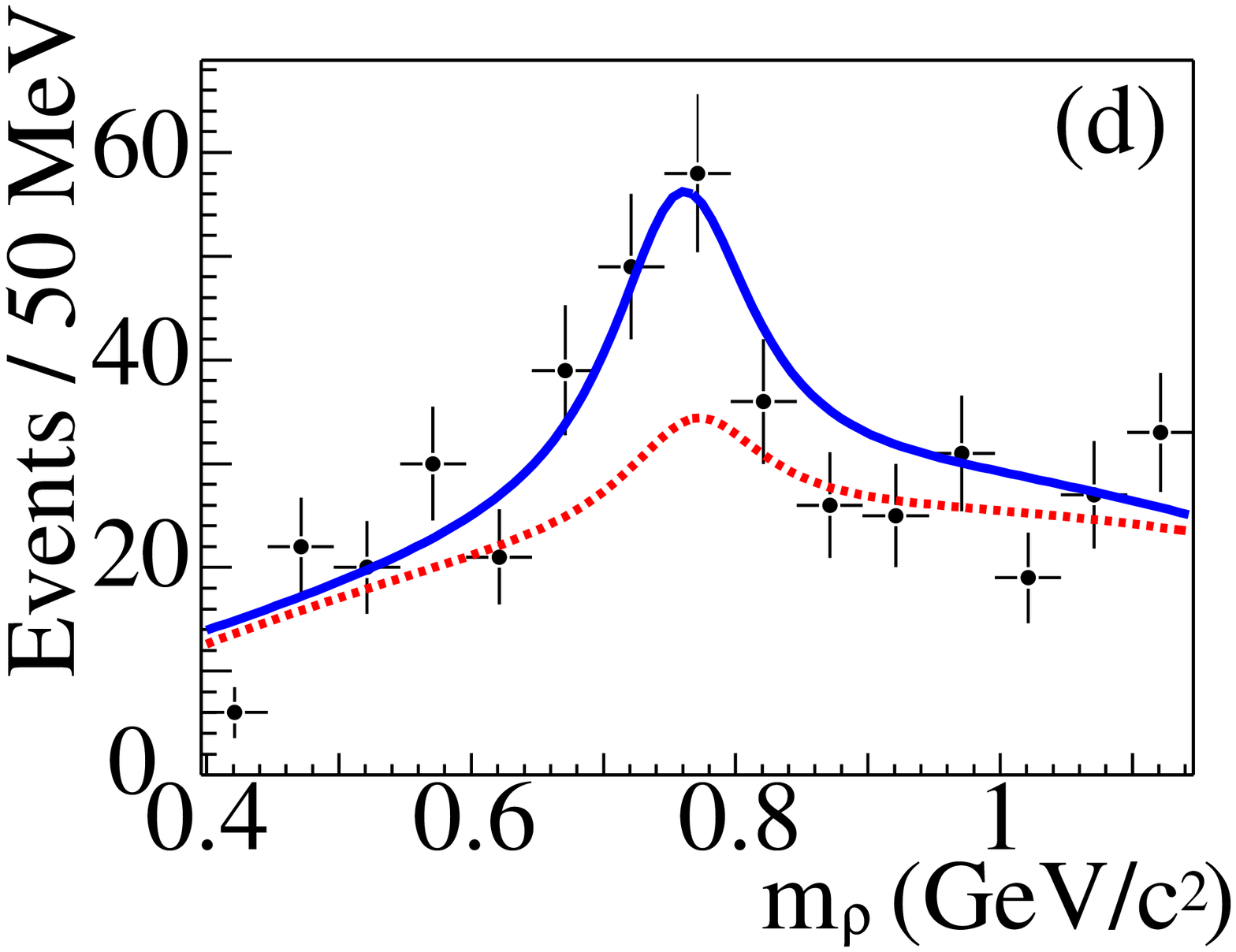}
}
\caption{The distributions for a sample of events enriched in signal for 
the variables (a) \mes, (b) \DeltaE, (c) cosine of the $\rho$ helicity 
angle and (d) \mv.  The dotted line is the projection of the sum of 
backgrounds and the solid line is the projection of the full likelihood.  
For \mes we show the projections for (top line) all and (bottom line) 
{\tt Lepton} and {\tt Kaon 1} tagged events.} \label{fig:plots}
\end{center}
\end{figure}

\begin{figure}[bp]
\begin{center}
\resizebox{6cm}{!}{
 \includegraphics{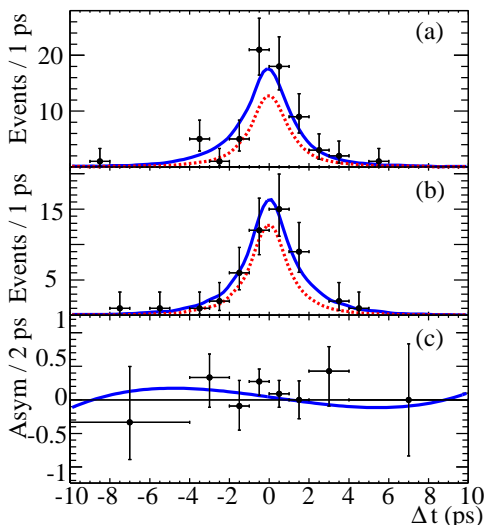}
} \caption{The \deltat\ distribution for a sample of events enriched in signal for (a) $\Bz$  and (b) $\Bzb$ tagged events.  The dashed line represents the sum of backgrounds and the solid line represents the sum of signal and backgrounds.  The time-dependent \CP asymmetry A (see text) is shown in (c), where the curve represents the asymmetry.} \label{fig:dtplots}
\end{center}
\end{figure}

We perform an unbinned extended maximum likelihood (ML) fit that assumes the event types
mentioned previously. The results of the fit are $246 \pm 29$ signal events with $\fL = 0.99 \pm
0.03$, $\slong = -0.42 \pm 0.42$ and $\clong = -0.17 \pm 0.27$.   There is a bias on the 
yield coming from the neglect of correlations in the fit (six events) and $B$-background
 modeling (16 events).  The former 
 is estimated using MC simulations and the latter is dominated by the change in signal yield 
when the $a_1\rho$ component is to allowed to vary in a fit to the data.  The signal yield 
remains stable when allowing the yield of other background types to vary. The corrected 
signal yield is $224 \pm 29$ events.
Figure~\ref{fig:plots} shows distributions of \mes, \coshel\ and \mv\ for 
{\tt Lepton} and {\tt Kaon 1} tagged events,
enhanced in signal content by cuts on the signal-to-background likelihood ratios of the 
discriminating variables not projected.  
The additional cuts retain ${\cal O} (15\%)$ of the signal events in the analysis sample.
For  \mes and \DeltaE\ we show a projection of the data for all tag categories; 
in these plots we retain ${\cal O} (60\%)$ of the 
signal events in the analysis sample.
Figure~\ref{fig:dtplots} shows the raw \deltat\ distribution for $\B^0$ 
and $\overline{\B}^0$ tagged events. The time-dependent decay-rate 
asymmetry $A = (
  R (\deltat) - \overline{R}(\deltat) ) / (  R (\deltat) + \overline{R}(\deltat) )$
is also shown, where $R$($\overline{R}$) is the decay-rate for \Bz(\Bzb) tagged events.

The nominal fit does not account for non-resonant background. If we add
a non-resonant component of $\B\to\rho \pi\pi^0$ events
to the likelihood, the fitted signal yield changes by less than $11$\% (90\% C.L.).  Any possible 
$\B\to4\pi$ component would be significantly smaller. 
The dominant systematic uncertainties in the yield arise from the assumed \B-background
branching fractions (20 events) and the uncertainty on the fraction of SCF events (14
events).  The uncertainty on the estimated fraction of misreconstructed events is
extrapolated from a control sample of fully reconstructed $B^{0} \rightarrow D^{-} \rho^{+}$
decays. 
A 10\% systematic error on the branching fraction comes from \piz reconstruction.
The dominant systematic error on $\ptrue$ is from the
uncertainty in PDF parameterization ($\pm 0.03$). 
We vary \CP-violation in the \B background within reasonable limits.  This is the main 
systematic uncertainty on the \CP results: 0.08 (0.11) on {\slong} (\clong). 
The systematic uncertainty on {\slong} (\clong) from B-background branching fractions 
is 0.02 (0.03).  Uncertainty in the vertex-detector alignment contributes an error of 
0.06 (0.04) on {\slong} (\clong). In half of the SCF events the misreconstructed signal 
contains at least one wrong track; the difference in  resolution function for these 
events corresponds to an uncertainty of 0.03 (0.01) on {\slong} (\clong).
The uncertainty in the parametrization of the likelihood contributes 
an error of 0.05 (0.02) on {\slong} ({\clong}).
We estimate the systematic error from ignoring interference with non-resonant modes and 
$a_1\pi$ to be 0.02 on \slong\ and \clong, and 2.4\% on the signal yield.
The uncertainty from possible \CP-violation in the doubly-Cabibbo-suppressed 
decays on the tag side of the event~\cite{ref:dcsd} is assumed to be
the same as for \Bztopippim: 0.012 (0.037) for {\slong} (\clong).  
We also apply a correction to account for possible dilution from \B-background 
(5\%) and SCF (3\%) events.

Our results are
\vspace{-0.9cm}
\begin{center}
\begin{eqnarray}
BR(B^0\to\rhoprhom) &=& (33 \pm 4 \stat \pm 5 \syst) \times 10^{-6} \nonumber\\
\ptrue&=& 0.99 \pm 0.03 \stat \;^{+0.04}_{-0.03} \syst \nonumber\\
\clong   &=& -0.17 \pm 0.27 \stat \pm 0.14 \syst \nonumber\\
{\slong} &=&  -0.42 \pm 0.42 \stat \pm 0.14 \syst \nonumber
\end{eqnarray}
\end{center}
The correlation coefficient between {\slong} and \clong\ is $-0.016$.  
We average this branching fraction with the less precise result from Ref. ~\cite{LBLrr}, 
taking into account correlations where appropriate~\cite{hfagrhorhozero}, to 
obtain the final value of $(30 \pm 4 \pm 5)\times 10^{-6}$.  This measurement supersedes the 
previous \babar\ result presented in Ref.~\cite{LBLrr}.

\begin{figure}[!htp]
\begin{center}
\resizebox{7cm}{!}{
 \includegraphics{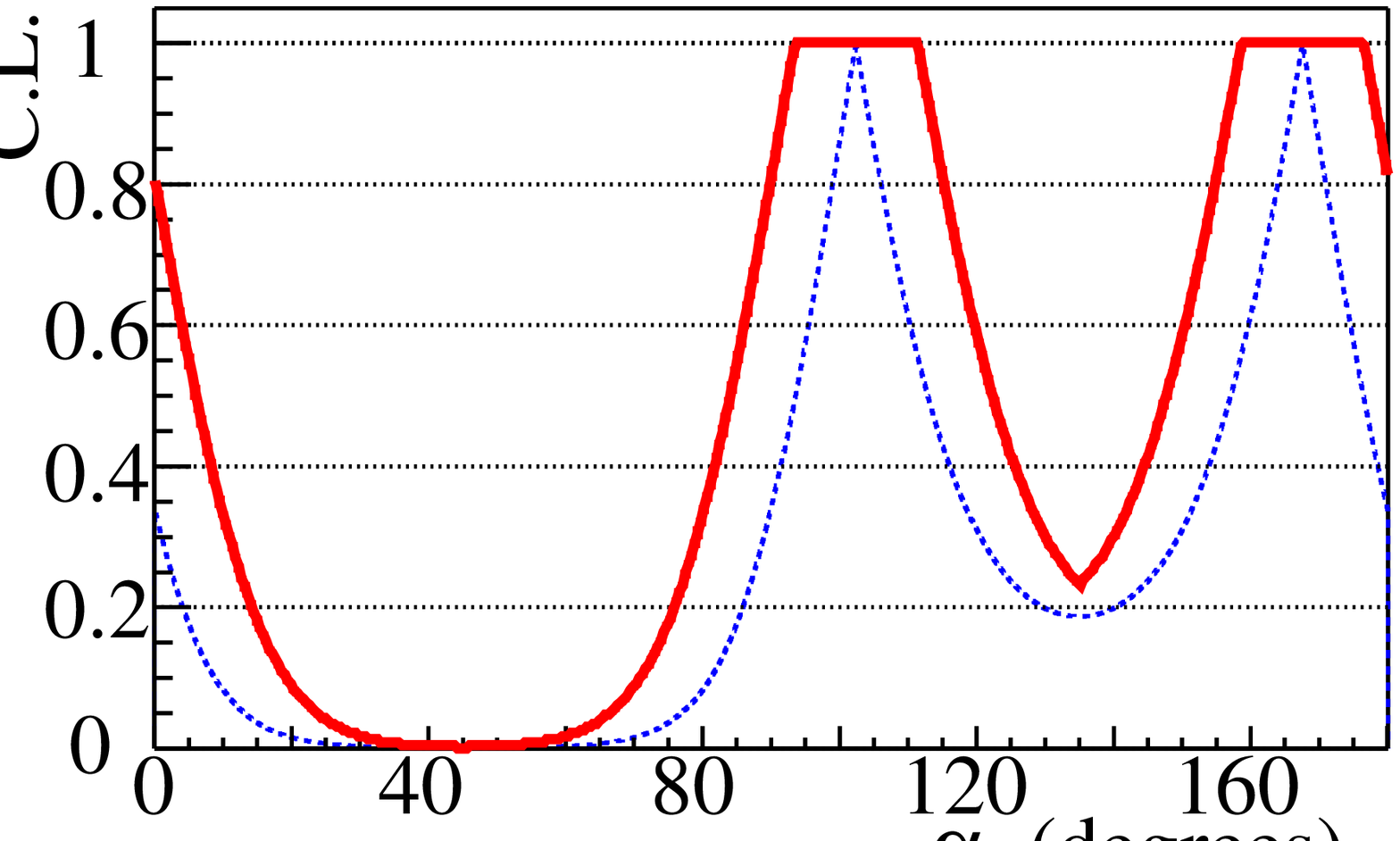}
} \caption{A plot of (dotted) $\alpha_{eff}$ and (solid) $\alpha$ as a function of confidence level for this result.} \label{fig:alpha}
\end{center}
\end{figure}

Using the Grossman-Quinn bound~\cite{grossmanquinn,LBLrr} with the recent results on $B\to\rho^\pm\rho^0, \rho^0\rho^0$ 
from~\cite{recentrhorho} we limit $|\alpha_{\mathrm{eff}}-\alpha|<13^\circ$ (68\% C.L.).
Ignoring possible non-resonant contributions, 
and $I=1$ amplitudes~\cite{falk} one can relate the \CP\ parameters {\slong} 
and \clong\ to $\alpha$, up to a four-fold ambiguity.
If we select the solution closest to the CKM best fit central value of 
$\alpha=95 - 98^\circ$~\cite{ckmbestfit}, the
 measured \CP\ parameters of the longitudinal polarization correspond to 
$\alpha = 102_{-12}^{+16} \stat ^{+5}_{-4} \syst \pm 13 ({\rm penguin})^\circ$.
Figure~\ref{fig:alpha} shows the confidence level as a function of 
$\alpha_{eff}={\rm arcsin}(\slong/\sqrt{1-\clong^2})/2$ for this result, (dotted) taking 
into account systematic uncertainties 
and (solid) also including the penguin contribution. We exclude values of 
$\alpha$ between $19^\circ$ and $71^\circ$ (90\% C.L.). 

We are grateful for the excellent luminosity and machine conditions
provided by our \pep2\ colleagues, 
and for the substantial dedicated effort from
the computing organizations that support \babar.
The collaborating institutions wish to thank 
SLAC for its support and kind hospitality. 
This work is supported by
DOE
and NSF (USA),
NSERC (Canada),
IHEP (China),
CEA and
CNRS-IN2P3
(France),
BMBF and DFG
(Germany),
INFN (Italy),
FOM (The Netherlands),
NFR (Norway),
MIST (Russia), and
PPARC (United Kingdom). 
Individuals have received support from CONACyT (Mexico), A.~P.~Sloan Foundation, 
Research Corporation,
and Alexander von Humboldt Foundation.


\end{document}